\documentclass[11pt,reqno]{amsart}

\usepackage{amssymb,amsmath}
\usepackage{graphicx}
\usepackage{bbm}
\usepackage{geometry}
\usepackage{float}
\usepackage{hyperref}
\usepackage{tikz}
\usetikzlibrary{matrix}
\usetikzlibrary{calc, shapes.geometric}
\tikzset{set/.style={draw,ellipse,inner sep=0pt,align=right, anchor=west}}
\usepackage[
  justification=RaggedRight,
  calcwidth=.9\linewidth,
]{caption}

\theoremstyle{plain}
\newtheorem{theorem}{Theorem}[section]
\newtheorem{proposition}[theorem]{Proposition}

\theoremstyle{definition}
\newtheorem{definition}[theorem]{Definition}
\newtheorem{remark}[theorem]{Remark}
\newtheorem{example}[theorem]{Example}
\newtheorem{question}[theorem]{Question}

\newcommand{\Z}{{\mathbb Z}}
\newcommand{\Q}{{\mathbb Q}}
\newcommand{\R}{{\mathbb R}}
\newcommand{\N}{{\mathbb N}}
\newcommand{\C}{{\mathbb C}}
\newcommand{\im}{{\mathrm{i}}}
\newcommand{\e}{\operatorname{e}}
\newcommand{\points}{\mathfrak{I}}

\newcommand{\eps}{\varepsilon}

\newcommand{\ts}{\hspace{0.5pt}}

\newcommand{\CC}{\mathbb{C}\ts}
\newcommand{\RR}{\mathbb{R}\ts}
\newcommand{\ZZ}{{\ts \mathbb{Z}}}

\newcommand{\NN}{\mathbb{N}}

\newcommand{\cL}{\mathcal{L}}

\newcommand{\dd}{\, \mathrm{d}}

\newcommand{\CalF}{\mathcal{F}}

\newcommand{\Hmm}[1]{\leavevmode{\marginpar{\tiny%
$\hbox to 0mm{\hspace*{-0.5mm}$\leftarrow$\hss}%
\vcenter{\vrule depth 0.1mm height 0.1mm width \the\marginparwidth}%
\hbox to 0mm{\hss$\rightarrow$\hspace*{-0.5mm}}$\\\relax\raggedright
#1}}}

\begin{document}

\title{Pure point diffraction and almost periodicity}

\author{Daniel Lenz}
\address{Mathematisches Institut,\newline 
\hspace*{\parindent}Friedrich Schiller Universit\"at Jena,\newline \hspace*{\parindent}07743 Jena, Germany}
\email{daniel.lenz@uni-jena.de}

\author{Timo Spindeler}
\address{Fakult\"at f\"ur Mathematik,\newline
\hspace*{\parindent}Universit\"at Bielefeld, \newline
\hspace*{\parindent}33501 Bielefeld, Germany}
\email{tspindel@math.uni-bielefeld.de}

\author{Nicolae Strungaru}
\address{Department of Mathematical Sciences,\newline 
\hspace*{\parindent}MacEwan University \newline
\hspace*{\parindent}10700 -- 104 Ave., Edmonton, AB, T5J 4S2, Canada \newline
\hspace*{\parindent} and  \newline
\hspace*{\parindent} Institute of Mathematics ``Simon Stoilow''  \newline
\hspace*{\parindent} Bucharest, Romania} 
\email{strungarun@macewan.ca}

\twocolumn[
\begin{@twocolumnfalse}
\begin{abstract}
This article deals with pure point
diffraction and its connection to various notions of almost periodicity.
We explain why the Fibonacci chain does not fit into the classical class of
Bohr almost periodicity and how it fits into the classes of mean, Besicovitch and Weyl almost periodic point sets. We report on recent results which characterize pure point diffraction as mean almost periodicity of the underlying structure,
and discuss how the complex amplitudes fit into this picture.
\end{abstract}
\maketitle
\end{@twocolumnfalse}
]


\section{Introduction}
The discovery of quasicrystals in 1982 by Dan Shechtman \cite{Danny} has led to a new area of mathematics, which is called aperiodic order. It comes with many mathematical questions, one of the most important
being: Which structures have pure point diffraction spectrum?
Accordingly, understanding pure point diffraction is a
central issue in the conceptual study of aperiodic order.

There is a strong connection between pure point diffraction
and almost periodicity. For a long time, this connection was little known,
but many hints and instances of it can be found, see for example \cite{ARMA,Sol1,Lag,BM,Gou-1,Meyer-generalized}.

There is a review article by Lagarias from 2000
\cite{Lag}, which states that a convincing framework
for the study of aperiodic order via almost periodicity should exist, and needs to be worked out.
Over the years, it has been observed that certain
notions of almost periodicity, such as Bohr (strong) or
weak almost periodicity, appear naturally in the study of the autocorrelation measure.
In this framework, pure point diffraction is equivalent to the strong almost periodicity of the
autocorrelation measure \cite{ARMA,MoSt}. However, as we will emphasize in Section~\ref{sect:fib}, for the point set itself,
these are not the appropriate notions.

In order to answer the question of which point sets have pure point diffraction, one thus needs to find a different
notion of almost periodicity. As we will explain below, this is possible.
Here, we will report on recent results by \cite{LSS} that develop
a theory of almost periodic measures that are centered around three
successively stronger instances of almost periodicity. Each of these
notions answers one important question in the  mathematical theory of pure point diffraction.

\smallskip

In order to illustrate the main point of this paper, we will first discuss the well-known Fibonacci chain. It can be defined via a substitution on letters, namely
\begin{equation} \label{eq:subsfibo}
\ell \mapsto \ell s \qquad \text{ and } \qquad s \mapsto \ell \ts,
\end{equation}
with the two letters $s$ and $\ell$, see \cite{gl} or \cite[Ch. 4]{TAO}. Starting from the legal pair of letters $\ell|\ell$ and applying \eqref{eq:subsfibo} over and over again, we obtain the words
\[
\ell s|\ell s,\ \ell s\ell|\ell s\ell,\ \ell s\ell\ell s|\ell s\ell\ell s,\  \ldots
\]
and, in the limit, the bi-infinite word
\[
\ldots \ell s\ell\ell s\ell s\ell\ell s\ell\ell s\ell s\ell\ell s\ell s\ell|\ell s\ell\ell s\ell s\ell\ell s\ell\ell s\ell s\ell\ell s\ell s\ell \ldots
\]
Next, we turn every $\ell$ into a long interval of length $\phi:=\frac{1+\sqrt{5}}{2}$ and every $s$ into a small interval of length $1$ to obtain a tiling of the real line. Alternatively,
this tiling can be constructed by starting with the intervals $\ell$ of length $\phi$ and $s$ of length $1$, and applying the geometric substitution from Figure~\ref{fig:geom_sub}, repeatedly.

\begin{figure}[H]
\begin{tikzpicture}[scale=.7]
\draw (-1,0)--(.61,0);
\draw (-1,-1)--(0,-1);
\draw (-1,-.1)--(-1,.1);
\draw (.61,-.1)--(.61,.1);
\draw (-1,-1.1)--(-1,-.9);
\draw (0,-1.1)--(0,-.9);
\draw (2, -.5)--(3,-.5);
\draw (2.7, -.7)--(3,-.5);
\draw (2.7, -.3)--(3,-.5);
\draw (4,0)--(6.61,0);
\draw (4,-.1)--(4,.1);
\draw (5.61,-.1)--(5.61,.1);
\draw (6.61,-.1)--(6.61,.1);
\draw (4,-1)--(5.61,-1);
\draw (4,-1.1)--(4,-.9);
\draw (5.61,-1.1)--(5.61,-.9);
\node[anchor=north] at (-.191,0){$\ell$};
\node[anchor=north] at (4.809,0){$\ell$};
\node[anchor=north] at (4.809,-1){$\ell$};
\node[anchor=north] at (-.5,-1){$s$};
\node[anchor=north] at (6.11,0){s};
\node[anchor=south] at (2.5,-.5){$\phi$};
\end{tikzpicture}
\caption{Geometric Fibonacci substitution, also called inflation.} \label{fig:geom_sub}
\end{figure}

 Note that the substitution inflates every tile by the factor $\phi$, see \cite{TAO} for more details. It is a well-established fact that the Fibonacci tiling gives rise to a diffraction spectrum that only shows bright spots --- a pure point diffraction spectrum. But, as we will explain in Section~\ref{sect:fib}, the Fibonacci chain is not almost periodic in the sense of Bohr!

\section{Almost periodic functions} \label{sec:almostper}

\begin{figure*}
\centering \includegraphics[scale=0.5]{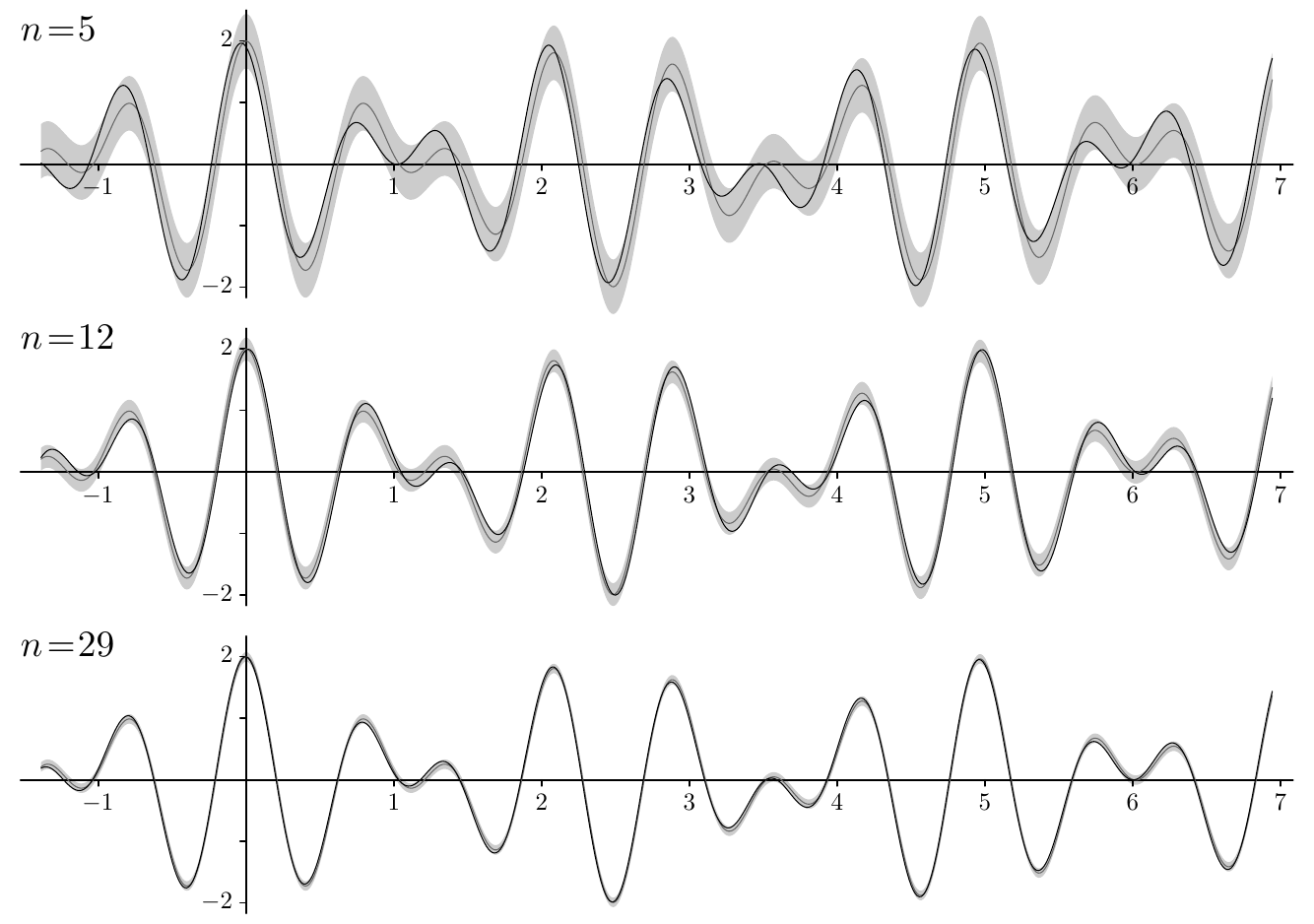}
\caption{A plot of the function $f(x)=\cos(2\pi x)+\cos(2\pi\sqrt{2}x)$ and  $f(x+n)$ for $n\in\{5,\, 12,\, 29\}$. Image taken from \cite{TAO} with permission.}\label{fig:bap}
\end{figure*}

To begin with, we will discuss the classical notions of periodic functions, quasiperiodic functions, limit-periodic functions, limit-quasiperiodic
functions, Bohr almost periodic functions and the relations among these. Let us start with the simple concept of a periodic function.

\begin{definition}
A function $f$ on $\R$ is called
\textit{periodic} if there is a number $L>0$ such that
\[
f(x+L)\, =\, f(x) \qquad \text{ for all } x\in\mathbb{R} \ts.
\]
\end{definition}
There are many well-known examples of periodic functions, such as
constant functions, $\sin(x)$, $\cos(x)$ or $\e^{\im x}$.

To every periodic function $f$, we can associate a \emph{Fourier series}
\smallskip
\begin{equation} \label{eq:fourierreihe}
  \sum_{m\in\mathbb{Z}} c_m\ts  \e^{2\pi\im \frac{m}{L}x}  \ts,
\end{equation}
where $c_m  =  \frac{1}{L} \int_0^L \e^{-2\pi\im\frac{m x}{L}}\,
f(x)\, \dd x$. Here, $L$ can be any (non-trivial) period of the function.

As shown by Carleson \cite{car}, if the function $f$ is nice, the Fourier
series of $f$ converges to $f(x)$ at almost every point $x$. On the other hand, there are
many periodic functions $f$ for which the Fourier series does not converge, see \cite{dbr}.

\smallskip

The sum of two periodic functions is in general not periodic, see
Figure~\ref{fig:bap}. However, if we shift the function by the right
amount, it looks almost like the unshifted version. This was
rigorously defined and developed by Harald Bohr, see \cite{bohr,bes}.

\begin{definition}
A continuous function $f:\R\to\C$ is called \textit{Bohr almost periodic} if, for all $\eps>0$, the set $P_{\eps}=\{t\in\R \,:\,
\|f-\tau_tf\|^{}_{\infty} <\eps\}$ of \textit{$\eps$-almost
periods} of $f$ is relatively dense in $\R$. Here, $\tau_tf(x) = f(x-t)$ denotes the translation of $f$ by $t$.
\end{definition}

Recall here that a subset $S$ of $\R$ is called \textit{relatively dense}
if there exists some interval $I$ such that, if we put a copy of $I$ at each point of $S$, we cover the entire real line. For example, the set
$\Z$ is relatively dense in $\R$, see \cite[p.~12]{TAO}.

Intuitively, a function $f$ is Bohr almost periodic if there are many $t$'s such that $\tau_t f$ is close to $f$.

\smallskip

The set of Bohr almost periodic functions can be characterized in many
different ways. One of the main theorems on Bohr almost periodic
functions reads as follows.

\begin{theorem}\cite[Prop. 8.2]{TAO} \label{thm:bohr}
A continuous function $f:\R\to \C$ is Bohr almost periodic if and only if
$f$ is the uniform limit of a sequence of trigonometric polynomials. \qed
\end{theorem}

Let us recall here that a trigonometric polynomial is a finite
sum of wave functions
\[
P(x)= \sum_{k=1}^N c_k\ts  \e^{2\pi\im y^{}_k x} \ts,
\]
where $c^{}_1,\ldots, c^{}_N \in \C$ and $y^{}_1, \ldots, y^{}_N \in \R$.

It is worth mentioning that the trigonometric approximations
are the key to undertstanding the connection between pure point diffraction and almost periodicity. This will be discussed in the following sections.

\smallskip

The above characterization suggests that Bohr almost periodic functions may be expandable in a way
similar to Eq.~\eqref{eq:fourierreihe}. As we do not have a common fundamental domain over which we can integrate, we instead consider the average integral
\[
A_k := \lim_{T\to\infty} \frac{1}{2T} \int_{-T}^T \e^{-2\pi\im kx}
f(x)\, \dd x \ts.
\]
$A_k$ is called the \emph{amplitude} (or the \emph{Fourier--Bohr coefficient}) of $f$ at the wave number $k$.

The \emph{Fourier--Bohr spectrum}
\[
\points:=\{k\in\R\,:\, A_k\neq 0\}
\]
is countable \cite[Thm.3.5]{bes}. The (formal)
\textit{Fourier--Bohr series} attached to $f$ is defined as
\[
\sum_{k\in \points} A_k\ts \e^{2\pi\im kx} \ts.
\]
Once again, the series need not converge to $f$, but it does converge
when $f$ is a nice function. Moreover, for periodic continuous functions,
the Fourier series and Fourier--Bohr series coincide \cite{bes}.

\smallskip

There are several important subsets of the set of Bohr almost periodic functions, which we recall below.
\begin{itemize}
\item \textit{Periodic functions}: We already discussed periodic functions earlier. They are exactly the Bohr almost periodic functions that satisfy $\points \subseteq \frac{2\pi}{L}\Z$, where $L>0$ is any period of $f$.

\item \textit{Limit-periodic functions}: A Bohr almost periodic function is limit-periodic if it is the (uniform) limit of a sequence of periodic functions. These functions are exactly the Bohr almost periodic functions that satisfy $\points \subseteq \frac{2\pi}{L}\Q$, for some $L>0$. It is important to note that the ratio of any two frequencies of such a function must be rational.

\item \textit{Quasiperiodic functions}: A Bohr almost periodic function is quasiperiodic when $\points$ can be indexed by finitely many fundamental frequencies.

\item \textit{Limit-quasiperiodic functions}:
A Bohr almost periodic function is limit-quasiperiodic when it is the (uniform) limit of a sequence of quasiperiodic functions.
\end{itemize}
Every periodic function is quasiperiodic and every quasiperiodic function is limit-quasiperiodic. Similarly,
every periodic function is a limit-periodic function, and every limit-periodic
function is also limit-quasiperiodic, but in a trivial way.

In order to get a better understanding of the differences between these kinds of almost periodic functions, let us take a look at some examples.

\begin{example}
 The function
\[
f(x) \,=\, \cos(2\pi x) + \cos(2\pi \sqrt{2} x)
\]
from Figure~\ref{fig:bap} is quasiperiodic, but neither periodic nor limit-periodic.

It is quasiperiodic since $f$ can
be written as
\[
f(x) \,=\, \frac{1}{2}\big( \e^{-2\pi\im\sqrt{2}x} \,+\,
\e^{-2\pi\im x} \,+\,  \e^{2\pi\im x} \,+\,
\e^{2\pi\im\sqrt{2}x} \big) \ts.
\] It cannot be periodic because
only $x=0$ satisfies the equation $f(x)=2$. It cannot be limit-periodic either because
the ratio of $\sqrt{2}$ and $1$ is not rational.
\end{example}

\begin{example}  \label{ex:sum_sine}
The function
\[
f(x) \,=\, \sum_{n\geqslant1} \frac{1}{n^2} \sin\big(2\pi\tfrac{1}{2^n}x\big)
\]
is limit-periodic but neither periodic nor quasiperiodic.
Note that
$\points= \big\{\pm\frac{1}{2^n} \,:\, n\in\N\big\}\subseteq \tfrac{2 \pi}{L}\Q$ for $L=2\pi$.
Hence, $f$ is limit-periodic but not periodic.
It is not
quasiperiodic either, since any finite set of fundamental frequencies only generates fractions with bounded denominator. Figure~\ref{fig:example} shows a plot of this function.

\begin{figure}[H]
\centering \includegraphics[scale=0.6]{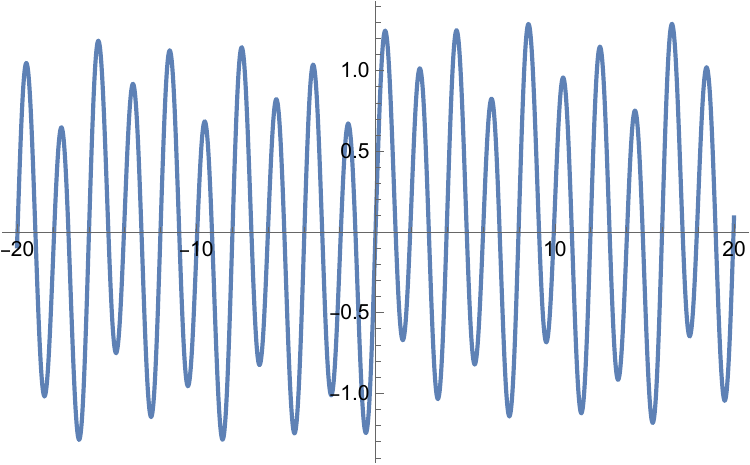}
\caption{A plot of the function $f$ from Example~\ref{ex:sum_sine}.} \label{fig:example}
\end{figure}

\begin{example} \label{ex:limquasi}
The function
\[
f(x) = \sum_{n\geqslant1} \frac{1}{n^3} \bigg( \sin\Big(\frac{2\pi x}{2^n}\Big) + \sin\Big(\frac{2\pi\sqrt{5} x}{2^n}\Big) \bigg)
\]
is obviously limit-quasiperiodic. But it cannot be limit-periodic, since the ratio of $1$ and $\sqrt{5}$ is not rational.
\end{example}

\begin{figure}[H]
\centering \includegraphics[scale=0.6]{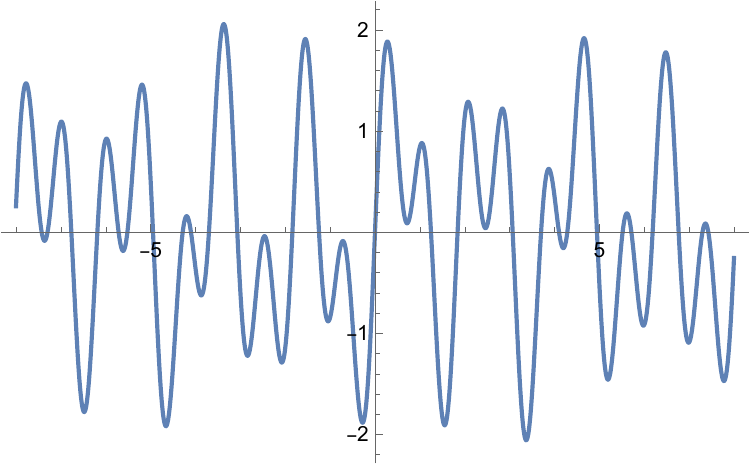}
\caption{A plot of the function $f$ from Example~\ref{ex:limquasi}.} \label{fig:example2}
\end{figure}

To summarize, we have the following hierarchy of almost periodic functions, and none of the implications is reversible:

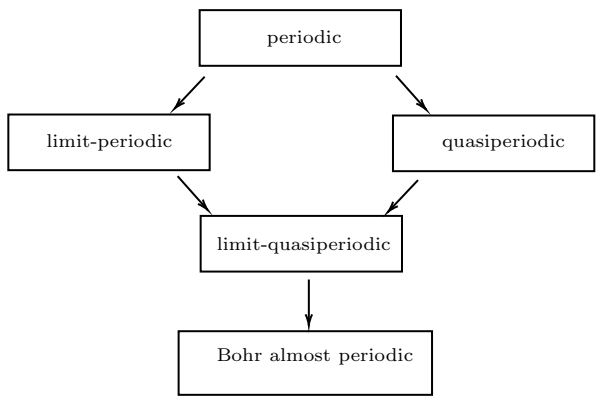
\begin{figure}[H]
\tikzset{every picture/.style={line width=0.75pt}} 
\begin{tikzpicture}[x=0.75pt,y=0.75pt,yscale=-0.5,xscale=0.5]
\draw   (231,23) -- (432,23) -- (432,79) -- (231,79) -- cycle ;
\draw   (424,129) -- (625,129) -- (625,185) -- (424,185) -- cycle ;
\draw   (40,128) -- (241,128) -- (241,184) -- (40,184) -- cycle ;
\draw   (232,230) -- (433,230) -- (433,286) -- (232,286) -- cycle ;
\draw   (210,346) -- (463,346) -- (463,410) -- (210,410) -- cycle ;
\draw    (236,90) -- (207.37,120.54) ;
\draw [shift={(206,122)}, rotate = 313.15] [color={rgb, 255:red, 0; green, 0; blue, 0 }  ][line width=0.75]    (10.93,-3.29) .. controls (6.95,-1.4) and (3.31,-0.3) .. (0,0) .. controls (3.31,0.3) and (6.95,1.4) .. (10.93,3.29)   ;
\draw    (449,194) -- (420.37,224.54) ;
\draw [shift={(419,226)}, rotate = 313.15] [color={rgb, 255:red, 0; green, 0; blue, 0 }  ][line width=0.75]    (10.93,-3.29) .. controls (6.95,-1.4) and (3.31,-0.3) .. (0,0) .. controls (3.31,0.3) and (6.95,1.4) .. (10.93,3.29)   ;
\draw    (427,89) -- (455.68,121.5) ;
\draw [shift={(457,123)}, rotate = 228.58] [color={rgb, 255:red, 0; green, 0; blue, 0 }  ][line width=0.75]    (10.93,-3.29) .. controls (6.95,-1.4) and (3.31,-0.3) .. (0,0) .. controls (3.31,0.3) and (6.95,1.4) .. (10.93,3.29)   ;
\draw    (209,190) -- (237.68,222.5) ;
\draw [shift={(239,224)}, rotate = 228.58] [color={rgb, 255:red, 0; green, 0; blue, 0 }  ][line width=0.75]    (10.93,-3.29) .. controls (6.95,-1.4) and (3.31,-0.3) .. (0,0) .. controls (3.31,0.3) and (6.95,1.4) .. (10.93,3.29)   ;
\draw    (340,294) -- (340,340) ;
\draw [shift={(340,340)}, rotate = 270] [color={rgb, 255:red, 0; green, 0; blue, 0 }  ][line width=0.75]    (10.93,-3.29) .. controls (6.95,-1.4) and (3.31,-0.3) .. (0,0) .. controls (3.31,0.3) and (6.95,1.4) .. (10.93,3.29)   ;
\draw (245,360) node [anchor=north west][inner sep=0.75pt]   [align=left] {{\tiny Bohr almost periodic}};
\draw (75,145) node [anchor=north west][inner sep=0.75pt]   [align=left] {{\tiny limit-periodic}};
\draw (470,145) node [anchor=north west][inner sep=0.75pt]   [align=left] {{\tiny quasiperiodic}};
\draw (295,40) node [anchor=north west][inner sep=0.75pt]   [align=left] {{\tiny periodic}};
\draw (245,248) node [anchor=north west][inner sep=0.75pt]   [align=left] {{\tiny limit-quasiperiodic}};
\end{tikzpicture}
\caption{Relations between the different kinds of almost periodicity. The arrows symbolize implications.}
\end{figure}
\end{example}

\section{The Fibonacci triangle function}\label{sect:fib}

\begin{figure*}
\centering
\begin{tikzpicture}[scale=.75]
\draw[->] (-10,0)--(10,0);
\clip (-10,-0.5) rectangle + (20,2.5);
\foreach \x in {-11.09, -8.472, -6.854, -4.236, -2.618, 0, 2.618, 4.236, 6.854, 9.472, 11.09, 13.71}
{
\draw[red] (0+\x,0)--(.809+\x,2)--(1.618+\x,0);
}
\foreach \y in {-9.472,-5.236,-1, 1.618 , 5.854 , 8.472, 12.708}
{
\draw[red] (0+\y,0)--(.5+\y,1)--(1+\y,0);
}
\end{tikzpicture}
\vspace*{-1.5cm}
\begin{tikzpicture}[scale=.75]
\draw[->] (-10,0)--(10,0);
\clip (-10,-2.5) rectangle + (20,5);
\foreach \x in {-11.09, -8.472, -6.854, -4.236, -2.618, 0, 2.618, 4.236, 6.854, 9.472, 11.09, 13.71}
{
\draw[red] (0+\x,0)--(.809+\x,2)--(1.618+\x,0);
\draw[blue] (-4.27+\x,0)--(.809+\x-4.27,2)--(1.618-4.27+\x,0);
}
\foreach \y in {-9.472,-5.236,-1, 1.618 , 5.854 , 8.472, 12.708}
{
\draw[red] (0+\y,0)--(.5+\y,1)--(1+\y,0);
\draw[blue] (-4.27+\y,0)--(.5+\y-4.27,1)--(1-4.27+\y,0);
}
\end{tikzpicture}
\caption{Top: Fibonacci triangle function. \\
 Bottom: Fibonacci triangle function $f$ ({\color{red}red}) and its translate $\tau_tf$ ({\color{blue}blue}) with $t=2 \phi+1$.} \label{fig:fibon}
\end{figure*}
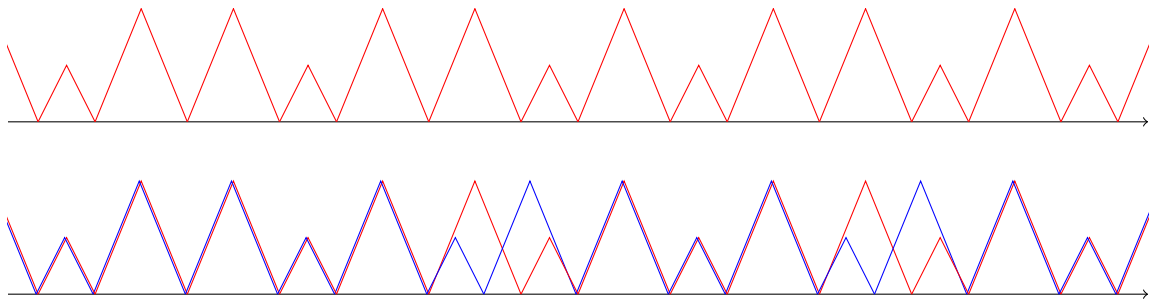

Let us discuss the basic issues around the various types of almost periodicity for model sets.

Consider the Fibonacci chain from the introduction. The set of all left end points of the corresponding intervals is a point set in $\R$, which we will denote by $\Lambda$. The distances between two consecutive
points can take only one of two values, namely $s=1$ and $\ell=\phi \approx 1.618$.
Let us put an isosceles triangle of height $1$ on each long tile $\ell$ and an isosceles triangle of height $\tfrac{1}{2}$ on each short tile $s$, see Figure~\ref{fig:fibon} (top). We claim that the function $f$ is not Bohr almost periodic.

Indeed, consider some $t$ such that the translate $\tau_t f$ is close to $f$.
Let us draw $f$ and $ \tau_t f$ on the same graph, see Figure~\ref{fig:fibon} (bottom). Now, $f(x)$ is zero at the end point of each tile of the Fibonacci chain, and hence $\tau_tf(x)$ is very close to $0$ at these points. This means that each point in the Fibonacci chain $\Lambda$ is close to some point in the translated Fibonacci chain $t+ \Lambda$. We can shift $t+\Lambda$ by a small $r$ such that some point in $\Lambda$ and some point in the new translate $r+t+\Lambda$ coincide. Note that, since $r$ is small, $\tau_{t+r} f$ is close to $f$.

Let us start at this common point and move to the left or right. Unless all points of $\Lambda$ and $r+t+\Lambda$ coincide, at some point, we hit a discrepancy. This means that we will hit a point which is the start of a pair of tiles in $\Lambda$ and $r+t+\Lambda$ which are different. We therefore have the following situation:

\begin{figure}[H]
\begin{tikzpicture}[scale=.5]
\draw[loosely dashed] (-7,0)--(2,0);
\draw[loosely dashed] (3.618,0)--(7,0);
\draw[red] (2,0)--(2+.809,2)--(2+1.618,0);
\draw[blue] (2,0)--(2.5,1)--(3,0);
\end{tikzpicture}
\end{figure}
\noindent The dashed line indicates that $f$ and $\tau_{t+r}f$ agree on this section of the plot, except for the two triangles, see also Figure~\ref{fig:fibon}. Since one of the triangles appears as part of the graph of $f$ and the other as a part of the graph of $\tau_{r+t}f$, the functions $\tau_{r+t} f$ and $f$ cannot be close in this situation.

The only way out is that $\Lambda$ and $r+t+\Lambda$ coincide, meaning $t+r=0$. It follows that $f$ cannot be Bohr almost periodic.

\medskip

It is not a coincidence that the Fibonacci triangle function is not Bohr almost periodic. Indeed, the Bohr/weak almost periodicity of a Delone set, combined
with the natural assumption of finite local complexity (FLC for short), implies full periodicity \cite{Fav,KL,LStru}. This means that among FLC Delone sets, Bohr almost periodicity identifies the fully periodic crystals. 

It is natural to ask if there are some notions weaker than Bohr almost periodicity which can identify aperiodic crystals. It turns out there are, and we will discuss them below. Let us first introduce them in the case of the Fibonacci chain.

\medskip

\begin{figure}[H]
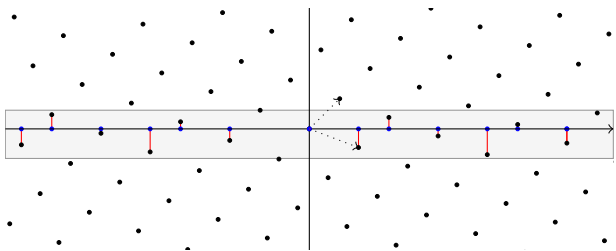

\tikz[scale=.4]{
\begin{scope}
\clip (-10,-4) rectangle + (20,8);
\draw[fill=gray!20,opacity=0.4] (-10,-.98)--(10, -.98)--(10, .619)--(-10,.619)--(-10,-.98);
\foreach \i\j in {0/0,1/0,1/1,2/1,3/1,3/2,4/2,4/3.-1/0,-1/-1, -2/-1,-2/-2,-3/-2,-4/-2,-4/-3}{
\draw[red] (\i*1.618+\j, -\i*.618+\j)--(\i*1.618+\j,0);
\node[draw,circle,inner sep=.5pt,fill,blue] at (\i*1.618+\j,0){};
}
\foreach \x in {-12,-11,...,12}{                           
    \foreach \y in {-12,-11,...,12}{                       
    \node[draw,circle,inner sep=.5pt,fill] at (\x*1.618+\y, -\x*.618+\y) {};}}
    \draw[->] (-10,0)--(10,0);
    \draw[->] (0,-8)--(0, 8);
\draw[->,dotted] (0,0)--(1,1);
\draw[->,dotted] (0,0)--(1.618,-.618);
\node[draw,circle,inner sep=.5pt,fill,blue] at (0,0){};
\draw[red] (0,0)--(0 ,0);
\end{scope}
}
\caption{The cut and project scheme of the Fibonacci chain.}\label{fig:fiboncps}
\end{figure}

Consider the Fibonacci chain as a model set, see Figure~\ref{fig:fiboncps}.
Let us recall that the Fibonacci model set is obtained by starting with the lattice
\[
\cL:= \ZZ \begin{pmatrix}
            1 \\
            1
          \end{pmatrix}+\ZZ \begin{pmatrix}
                              \phi \\
                              \phi'
                            \end{pmatrix}= \left\{ \begin{pmatrix}
            m+n \phi \\
            m+n \phi'
          \end{pmatrix}: m,n \in \ZZ \right\}
\]
in $\RR^2$ and projecting all the points in $\cL$ whose second coordinate lies in the interval $[-1, \phi-1)$ onto the first copy of $\RR$, see \cite[Ex.~7.3]{TAO} for more details. Here, $\phi'=\frac{1-\sqrt{5}}{2} \approx -0.618$ is the algebraic conjugate of $\phi$.

Now, for all $(t,r) \in \cL$ such that $r$ is close to $0$, the difference between the Fibonacci chain $\Lambda$ and its translate $t+ \Lambda$ by $t$ is the model set given by the difference between the window $W$ and its translate $r+W$. Since $r$ is small, $W$ and $r+W$ mostly overlap. This implies that, on average, $\Lambda$ and $t+ \Lambda$ ``almost agree". It follows that, for every $(t,r) \in \cL$ such that $r$ is close to $0$, the functions $f$ and $\tau_t f$ ``almost agree" on average, meaning that
\[
\lim_{n\to\infty} \frac{1}{2n} \int_{-n}^n | f(x)- \tau_t f(x)|\, \dd x
\]
is very small.

This shows that the set of $t \in \RR$ for which $f$ and $\tau_t f$ almost agree on average, is relatively dense. We will refer to this property as almost periodicity in mean (or average), and we will simply say that $f$ is \textit{mean almost periodic}. For a precise definition of mean almost periodic functions, we refer the reader to Definition~\ref{def:map} in Appendix A.

\medskip

In fact, the Fibonacci triangle function satisfies a property which is stronger than mean almost periodicity: on (uniform) \textit{average}, we can approximate $f$ as well as we want by trigonometric polynomials. This can be shown by approximating the window of the cut and project scheme by continuous functions. Since the details are technical, we skip them and refer the reader to \cite{LSS}, see also \cite{CR,Meyer-generalized,Nicu}. Below, we will refer to any function which, on average, can be approximated by trigonometric polynomials as a \textit{Besicovitch almost periodic function}, and to any function which, on average, can be approximated uniformly by trigonometric polynomials as a \textit{Weyl almost periodic function}, see Definitions~\ref{def:Bap} and \ref{def:wap} in the Appendix for the precise definitions.

Our goal in this paper is to introduce the reader to mean, Besicovitch and Weyl almost periodic functions and their relevance to pure point diffraction.
We will skip the technical details, and refer the reader to \cite{LSS} instead.

\begin{figure*}
\begin{tabular}{ccc}
\includegraphics[scale=.325]{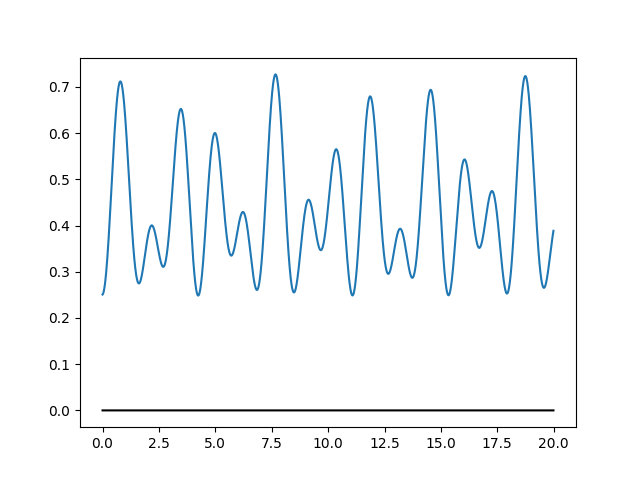} &
\includegraphics[scale=.325]{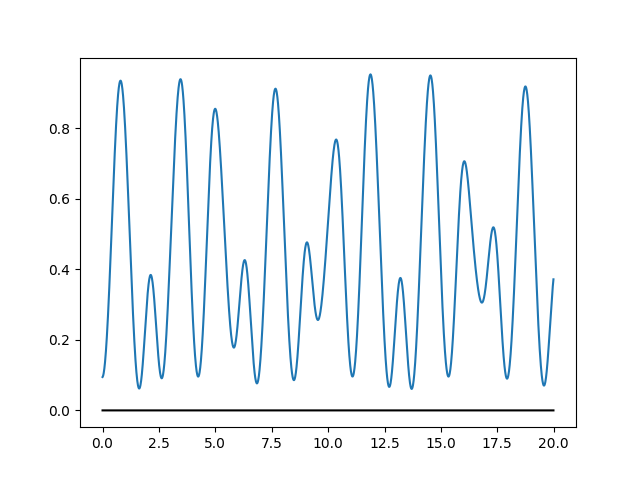} &
\includegraphics[scale=.325]{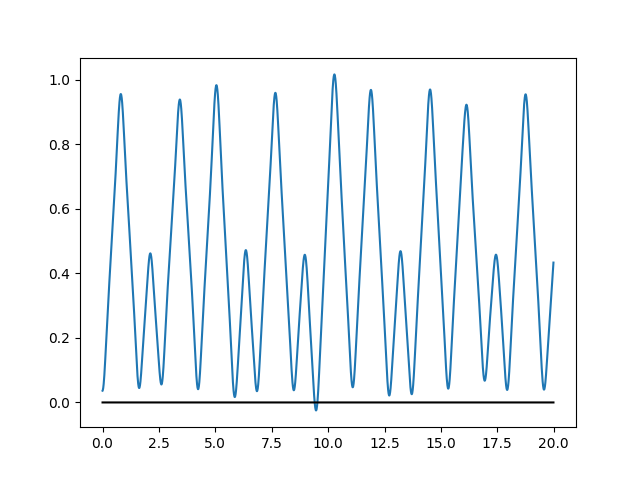} 
\end{tabular}
\caption{Three partial Fourier--Bohr series for the Fibonacci triangle function of Figure~\ref{fig:fibon}.}
\end{figure*}

\section{Pure point diffraction}

Below, we use the setup of mathematical diffraction theory.
The formal mathematical definitions and references can be found in Section~\ref{section-setup-diff} in the Appendix.

One of the most basic and fundamental questions in diffraction theory is
which point sets $\Lambda$ in $\RR^d$ have pure point diffraction. The answer is given by the following
result.

\begin{theorem}[{{\cite[Thm.~2.13]{LSS}}{ Characterizing pure point diffraction}}]
 A Delone set $\Lambda$ has pure point spectrum if and only if, for all compactly supported continuous functions $\varphi$, the function
 \begin{equation}\label{eqNphi}
 N_{\varphi}(x):= \sum_{y \in \Lambda} \varphi(x-y)= (\delta_{\Lambda}*\varphi)(x) 
 \end{equation}
 is mean almost periodic. \qed
\end{theorem}
Note that $N_{\varphi}$ is simply the function obtained by putting a copy of $\varphi$ at each point of $\Lambda$ and adding everything up.
For simplicity, whenever $ N_{\varphi}$ is a mean almost periodic function, for all test functions $\varphi$, we will simply say that $\Lambda$ is a \textit{mean almost periodic point set}.
Intuitively, this means that pure point diffraction happens exactly when there are many translates $t \in \RR^d$ such that, on average, $t+\Lambda$ and $\Lambda$ almost agree.

As we have seen above, the Fibonacci model set has this property. On the other hand, given a random structure $\Lambda \subset \RR^d$ and some $t \neq 0$, we should not expect much agreement between $\Lambda$ and $t+\Lambda$, and hence, the diffraction spectrum will contain a non-trivial continuous component.

\medskip

Next, we will discuss the so called \textit{consistent phase property} (or CPP for short). For homogeneous point sets $\Lambda \subset \RR^d$, the intensity of the Bragg peak at position $y \in \RR^d$ is given by the absolute value square of the (complex) \emph{amplitude}
\[
A_y:= \lim_{n\to\infty} \frac{1}{(2n)^d} \sum_{x \in \Lambda \cap [-n,n]^d} \e^{- 2 \pi \im x \cdot y}
\]
(where $x\cdot y$ denotes the inner product of $x$ and $y$), that is
\begin{equation}\label{eq1}
I(y)= \left| A_y \right|^2 \ts.
\end{equation}
Let us emphasize here that the CPP does not always
hold. There are non-homogeneous systems for which the limit $A_y$ does not always exist, see Example~\ref{ex2} below.
Also, it is possible that $A_y$ exists but \eqref{eq1} fails, see Example~\ref{ex1}.

We say that a point set $\Lambda \subset \RR^d$ satisfies the CPP if \eqref{eq1} holds for all $y \in \RR^d$. This property is an important one for physical models.
Whenever the CPP holds, one can recover the absolute value of $A_y$ from the intensity of Bragg peak. If one can further find out the phase information for $A_y$ (which is, in general, a difficult task), and the system has pure point diffraction, then Theorem~\ref{thm-RF} can be used to reconstruct $\Lambda$.

It is therefore a natural question to ask which systems satisfy the CPP. For systems with pure point spectrum, the answer is by the following.

\begin{theorem}[{{\cite[Thm. 3.36]{LSS}}{ Characterizing pure point diffraction and CPP}}]
 A Delone set $\Lambda$ has pure point spectrum and satisfies the CPP if and only if, for all compactly supported continuous functions $\varphi$, the function $N_\varphi$ of \eqref{eqNphi} is Besicovitch almost periodic. \qed
\end{theorem}

For simplicity, whenever $ N_{\varphi}$ is a Besicovitch almost periodic function, for all test functions $\varphi$, we will simply say that $\Lambda$ is a \textit{Besicovitch almost periodic point set}.

\smallskip

Let us now discuss some examples.

\begin{example}
Let $\Lambda$ be the Fibonacci model set. Then, $\Lambda$ is a Besicovitch almost periodic point set, compare \cite[Thm.~4.26]{LSS}. Therefore,
$\Lambda$ has pure point diffraction and satisfies the CPP.

Section~\ref{sect:fib} shows that $\Lambda$ is not Bohr almost periodic. We will discuss the Fourier expansion of $\Lambda$ in Theorem~\ref{thm-RF} below.
\end{example}

\begin{example}\label{ex-sf} Let
 \begin{align*}
\Lambda & =\Z\setminus\bigcup_{p \text{ prime}} p^2\Z \\
     &= \{0,\pm1,\pm2,\pm3,\pm5,\pm6,\pm7,
        \pm10, \pm11,\ldots \}
\end{align*}
be the set of square-free integers, that is, $\Lambda$ consists of all integers that are not divisible by the square of any prime. This set is highly ordered, but also contains larger and larger holes at sparse locations, see for example \cite{BMP,HR,BHS}.

$\Lambda$ is a Besicovitch almost periodic point set, see for example \cite[Prop.~3.39]{LSS}. Therefore, it has pure point diffraction and satisfies the CPP.
The diffraction of $\Lambda$ is drawn in Figure~\ref{fig:example_sfi} below.
\end{example}

\begin{figure}[H]
\centering \includegraphics[scale=0.5]{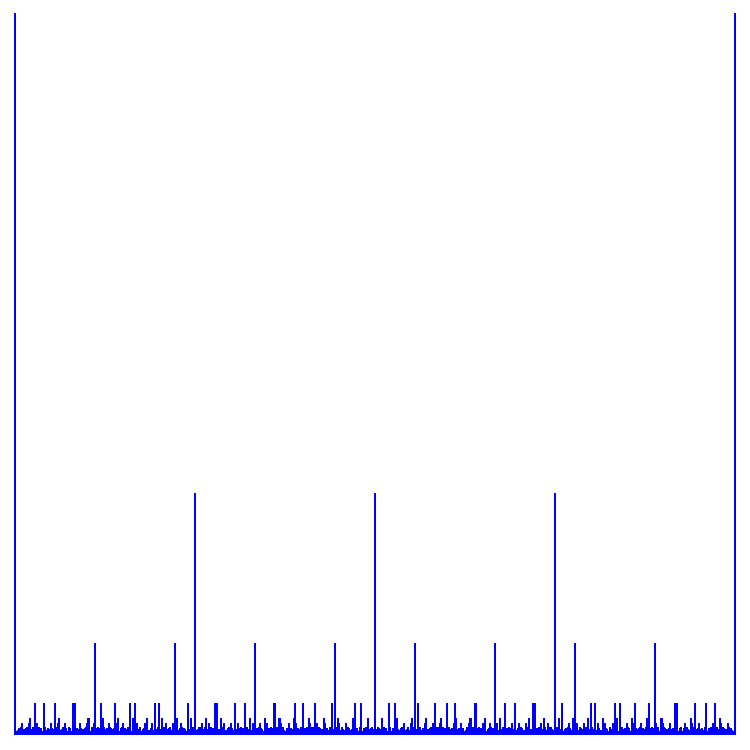}
\caption{A plot of the $1$-periodic diffraction pattern of the square-free integers. The plot shows the square root of the actual intensities on the interval $[0,1]$.}\label{fig:example_sfi}
\end{figure}

\begin{example}[Mean but not Besicovitch almost periodic]\label{ex1}
Consider the point set
\begin{align*}
\Lambda &= \{ -n \,:\, n \in \NN \}
 \cup \Big\{ \frac{1}{2\sqrt{2}} +n \,:\, n \in \NN \Big\} \ts,
\end{align*}
which consists of all negative integers and all positive integers, which are shifted to the right by $\tfrac{1}{2\sqrt{2}}$.

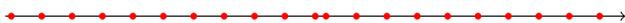
\begin{figure}[H]
\begin{tikzpicture}[scale=.4]
\begin{scope}
\draw[->] (-10.2,0)--(10.2, 0);
\foreach \x in {0,1,...,9}{                           
    \node[draw,red,circle,inner sep=.75pt,fill] at (-\x,0) {}; 
    \node[draw,red,circle,inner sep=.75pt,fill] at (\x+0.353,0) {};
    }
\node[draw,red,circle,inner sep=.75pt,fill] at (-10,0) {};
\end{scope}
\end{tikzpicture}
\caption{Part of the point set $\Lambda$ from Example~\ref{ex1}.}
\end{figure}

It is easy to see that the diffraction of $\Lambda$ is pure point, consisting of a Bragg peaks at each integer with the same intensity $I(m)=1$. It follows that
$\Lambda$ is a mean almost periodic point set.

On the other hand, for each $y \in \RR$, the amplitudes can be calculated explicitly, see \cite[Append. A2]{LSS} for details. It is non-zero only at the integers, and its value at $m \in \ZZ$ is
\[
A_m= \frac{1+\e^{2 \pi \im m \sqrt{2}}}{2} \ts.
\]
In particular, for all $m \in \ZZ\setminus\{0\}$, we have
\[
I(m) \neq | A_m|^2 \,.
\]
Since the CPP fails, $\Lambda$ is not Besicovitch almost periodic.

It is interesting that this point set has exactly the same diffraction as $\ZZ$. As it does not satisfy the CPP, it cannot be recovered from its diffraction.
This point set is not homogeneous, which has to be the case for any point set which is mean almost periodic, but not Besicovitch almost periodic.
\end{example}

Let us next see another example of a point set which is mean but not Besicovitch almost periodic. As the details are more technical, we skip them.

\begin{example}\label{ex2} Let
\begin{align*}
\Lambda &= \left\{ m \in \ZZ \,:\, \exists\, n \in \NN \mbox{ s.t. } 2^{2n} \leqslant |m| < 2^{2n+1} \right\} \ts.
\end{align*}
The set $\Lambda$ consists of all integers whose binary representation has an odd number of digits.

\begin{figure}[H]
\begin{tikzpicture}[scale=.11]
\begin{scope}
\draw[->] (-36.2,0)--(36.2, 0);
\foreach \x in {1,4,5,6,7,16,17,18,19,20,21,22,23,24,25,26,27,28,29,30,31}{                           
    \node[draw,red,circle,inner sep=.65pt,fill] at (-\x,0) {};
        \node[draw,red,circle,inner sep=.65pt,fill] at (\x,0) {};
    }
\end{scope}
\end{tikzpicture}
\caption{Part of the point set $\Lambda$ from Example~\ref{ex2}.}
\end{figure}
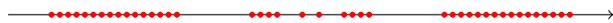

A simple but technical computation shows that $\Lambda$ is a mean almost periodic point set. On the other hand, the amplitude $A_m$ does not exist for each $m \in \ZZ$, so $\Lambda$ is not Besicovitch almost periodic.
\end{example}

\medskip

Another important property, which we are interested in, is the Fourier expansion of an almost periodic point set. By this, we mean that the equality
\[
\delta_{\Lambda} = \sum_{y \in \RR^d} c_y \ts\e^{2 \pi \im y \cdot(.)}
\]
holds in a certain sense, see Theorem~\ref{thm-RF}.
The existence of the Fourier expansion is covered by the following result.

\begin{theorem}[{{\label{thm-RF}\cite{LSS}}{ Fourier expansion}}]
A point set $\Lambda$ is Besicovitch almost periodic if and only if it has a  Fourier expansion in the following sense: for all compactly supported continuous functions $\varphi$, the function $N_\varphi$ of \eqref{eqNphi} satisfies the identity
\begin{equation}\label{eq-RF}
N_\varphi(x)= \sum_{y \in \RR^d} A_y \, \widehat{\varphi}(y)  \e^{2 \pi \im y \cdot x }
\end{equation}
on average. \qed
\end{theorem}

Note that $\widehat{\varphi}$ or $\mathcal{F}\varphi$ denotes the Fourier transform of $\varphi$, which is given by
\[
\widehat{\varphi}(x) :=  (\mathcal{F}\varphi)(x) := \int_{\R^d} \e^{-2\pi\im x\cdot y}\, \varphi(y)\, \dd y \ts.
\]

As we have seen above, the Fibonacci point set is Besicovitch almost periodic and hence satisfies \eqref{eq-RF}. As it is not Bohr almost periodic, the sum on the right hand side of the equality does \emph{not} converge uniformly to $N_\varphi$, but only on avarage.

\section{Some extensions}
In this section, we discuss two extensions of the above results. The first one is from point sets $\Lambda$ to measures, while the second replaces the averaging sequence $[-n,n]^d$ by more general ones.

\subsection{Extension to measures} There is an alternative approach to modeling mathematical diffraction: by using densities instead of point sets.
In this section, we review the mathematical framework of measures, which unifies the two seemingly incompatible approaches.
To do this, we will work with so called \textit{(translation-bounded) measures} on $\RR^d$, see \cite{BL-1} or \cite[Sect.~8.5]{TAO} for definitions and properties.

We refrain from giving the precise definition of a measure (see the quoted
literature), and emphasize instead that the characteristic feature of a measure
$\mu$  on $\RR^d$ is that we can talk about the convolution $\mu \ast
\varphi$ of $\mu$ with any function $\varphi \in C_c (\RR^d)$
(continuous functions on $\R^d$ with compact support). The function
\[
N_{\varphi}= \mu*\varphi 
\]
will always be continuous and bounded on $\RR^d$.

As above, a point set $\Lambda$ gives rise to its \textit{Dirac comb}
$\delta_\Lambda =\sum_{x \in \Lambda} \delta_x$, which is a measure, with
\[
N_{\varphi} (x)  = \sum_{y\in\Lambda}
\varphi(x-y) \ts,
\]
which is exactly the function from Eq.\eqref{eqNphi}.

Similarly, a model with a density function $f$ can be interpreted as a measure
via
\[
N_{\varphi}(x) = \int_{\RR^d} \varphi (x-y) f(y)\, \dd s \ts.
\]

\smallskip

We can now talk about almost periodic measures. Similar to the previous section, we will say that $\mu$ is a
mean, Besicovitch or Bohr almost periodic measure\footnote{Note that, in the mathematics literature, a Bohr almost periodic measure is usually called strongly almost periodic.} if the function $N_\varphi$
is mean, Besicovitch or Bohr almost periodic, for all compactly supported
continuous functions $\varphi$.

\medskip

Exactly like for point sets, we can ask the following questions.

\begin{question}
\begin{itemize}
  \item[(1)] Which measures $\mu$ have a pure point diffraction spectrum?
  \item[(2)] When does the CPP hold? More precisely, for which measures $\mu$ is the intensity of Bragg peaks $I(y)$ given by $I(y)= |A_y|^2$, where
\[
A_y:= \lim_{n\to\infty} \frac{1}{(2n)^d}  \int_{\RR^d} \e^{- 2 \pi \im x \cdot y}\, \dd \mu(y) \ts?
\]
  \item[(3)] Which measures $\mu$ have a Fourier expansion of the form
\[
\mu = \sum_{y \in \RR^d} A_y \, \e^{2 \pi \im y\cdot(.)} \ts,
\]
and in which sense does this hold?
\end{itemize}
\end{question}

The answer to these questions is similar to the one for point sets, and reads as follows.

\bigskip

\begin{theorem}[{\cite{LSS}}] \phantom{X}
\begin{itemize}
  \item[(1)] A measure $\mu$ has pure point diffraction if and only if $\mu$ is a mean almost periodic measure.
  \item[(2)] A measure $\mu$ has pure point diffraction and satisfies the CPP if and only if $\mu$ is a Besicovitch almost periodic measure.
  \item[(3)] A measure $\mu$ has a Fourier expansion of the form
\[
\mu = \sum_{y \in \RR^d} A_y \, \e^{2 \pi \im y \cdot (.)} \,
\]
in the sense that, for all compactly supported continuous functions $\varphi$, the equality
\[
N_{\varphi}(x)= \sum_{y \in \RR^d} A_y \, \widehat{\varphi}(y) \e^{2 \pi \im y\cdot x}
\]
holds on average if and only if $\mu$ is a Besicovitch almost periodic measure.  \qed
\end{itemize}
\end{theorem}

\subsection{Other averaging sequences}

In the previous sections, we have always averaged using the sequence $(A_n)$ with $A_n = [-n,n]^d \subset \RR^d$. One could instead use translates of these cubes, or balls or even more general van Hove sequences, see \cite[Def.~2.9]{TAO} for the formal definition.

When going to this generality, it becomes natural to ask how changing the averaging sequence will affect the diffraction measure. By allowing for translates of our cubes, balls, or more general averaging sequences, we are
studying if and how picking samples from different areas of our point sets changes the diffraction. In other words, we are looking at the homogeneity (or the lack thereof) of our point set. In particular, we would like to answer the following question.

\begin{question}
Which point sets $\Lambda$ (or, more generally, measures) have the property that, for all van Hove sequences, the diffraction is the same pure point measure and that the CPP holds?
\end{question}

Intuitively, we are asking which structures are sufficiently homogeneous and have a pure point diffraction measure.

To answer this question, we need to introduce a concept which is stronger than (but similar to) Besicovitch almost periodicity: Weyl almost periodicity.
A function $f$ is called \textit{Weyl almost periodic} if it can be approximated by trigonometric polynomials $P(x)= \sum_{k=1}^m c_k \e^{2 \pi \im y_k\cdot  x}$ in
the uniform average
\[
\lim_{n\to\infty}  \sup_{t\in\R^d} \int_{[-n,n]^d} \frac{|f(x-t)-P(x-t) |}{(2n)^d}\, \dd x \ts.
\]
Intuitively, Weyl almost periodicity requires not only $f$, but also all translates of $f$ to be approximated in average by the corresponding translates of the same trigonometric polynomial $P$, in a uniform way.

Now, a  measure $\mu$ is called Weyl almost periodic if the function $N_\varphi$ is Weyl almost periodic, for all compactly supported continuous functions $\varphi$.

One has the following characterization for uniform pure point diffraction and CPP.

\begin{theorem}[{{\cite{LSS}}{ Independence of the averaging sequence}}] A measure $\mu$ is Weyl almost periodic if and only if the following three conditions hold:
\begin{itemize}
  \item The diffraction measure for $\mu$ is independent of the choice of the van Hove sequence.
  \item The (complex) amplitudes are independent of the choice of the van Hove sequence.
  \item The diffraction measure for $\mu$ is pure point and satisfies the CPP.  \qed
\end{itemize}
\end{theorem}

The Fibonacci set, as well as any regular model set, is  Weyl almost periodic \cite{LSS}. Any Weyl almost periodic measure is automatically Besicovitch almost periodic.
The square free integers from Example~\ref{ex-sf} are Besicovitch almost periodic but have larger and larger holes, so they cannot be Weyl almost periodic.

The following hierarchy of almost periodic functions carries over to measures.

\medskip

\begin{tikzpicture}[scale=.4]
\node[set,text width=4.9cm, minimum height=2.7cm] {mean};
\node[set,text width=3.3cm, minimum height=2.1cm] {Besi-\\ covitch};
\node[set,text width=1.8cm, minimum height=1.5cm]  {Weyl};
\node[set,text width=.8cm, minimum height=.6cm] {Bohr};
\end{tikzpicture}

Lastly, let us discuss the Fourier expansion of Weyl almost periodic measures. Since any such measure is Besicovitch almost periodic, it has a Fourier expansion in the sense that
\begin{equation}\label{eq2}
N_{\varphi}(x)= \sum_{y \in \RR^d} c_y \, \widehat{\varphi}(y) \e^{2 \pi \im y\cdot x}
\end{equation}
holds on average. In fact, a measure is Weyl almost periodic exactly when \eqref{eq2} holds in the uniform average \cite{LSS}.

Let us emphasize that, for aperiodic crystals with finite local complexity, the equality in Eq.~\eqref{eq2} cannot hold in the sense
of uniform convergence, as this would imply Bohr almost periodicity and hence full periodicity. It can only hold on (in the uniform) average.

\section{Summary}

Our main results can be phrased as follows:
\begin{itemize}
  \item A point set or a measure is pure point diffractive if and only if it is mean almost
periodic.
  \item A point set or a measure is pure point diffractive and satisfies the CPP if and only if it is Besicovitch almost periodic.
  \item A point set or a measure is pure point diffractive, the diffraction is independent of the choice of the van Hove sequence and satisfies the CPP
if and only if it is Weyl almost periodic.
  \item Besicovitch (Weyl) almost periodic measures have Fourier expansions, which hold (uniformly) on average.
\end{itemize}

\appendix

 \section{Mean, Besicovitch and Weyl almost periodic functions}

Here, we recall the concepts for the notions of almost
periodicity we introduced in this review, compare \cite{bes,LSS}. For simplicity, we will introduce the ideas when $d=1$, and simply state that the general case is analogous.

Let us start by defining the \emph{Besicovitch semi-norm} $\|f\|^{}_{\mathcal{B}}$ of a bounded function $f: \RR \to \CC$ as
\[
\|f\|^{}_{\mathcal{B}}:=\limsup_{n\to\infty} \frac{1}{2n} \int_{-n}^n
|f(x)|\, \dd x \ts.
\]
Note that $\|\cdot\|^{}_{\mathcal{B}}$ is not a norm
but a semi-norm. This means that there are functions $f\neq0$ with
$\|f\|^{}_{\mathcal{B}}=0$, for example continuous functions with compact support.

We can now define mean and Besicovitch almost periodicity.

\begin{definition}\label{def:map}
A continuous function $f:\R\to\C$ is called \textit{mean almost
periodic} if, for each $\eps>0$, the set
\[
\{t\in \R \,:\, \|\tau_tf-f\|^{}_{\mathcal{B}}<\eps\}
\]
is relatively dense.
\end{definition}

This definition resembles the definition of Bohr almost periodic functions. The difference is that we use the Besicovitch semi-norm instead of the supremum norm.

\begin{remark}
 Bohr almost periodicity implies mean almost periodicity, which immediately follows from $\|f\|^{}_{\mathcal{B}}\leqslant\|f\|^{}_{\infty}$.
In particular, every periodic, limit-periodic, quasiperiodic and limit-quasiperiodic function is mean almost periodic.

 The function from Figure~\ref{fig:fibon} is mean almost periodic, but not Bohr almost periodic as we shown in Section~\ref{sect:fib}.
\end{remark}

Theorem~\ref{thm:bohr} showed that Bohr almost periodic functions
can be characterized either by relatively dense sets of
$\eps$-almost periods or by uniform approximation by trigonometric
polynomials. It is only natural to ask if the same is true for the
semi-norm $\|\cdot\|^{}_{\mathcal{B}}$.

Let us first introduce the following definition.

\begin{definition}\label{def:Bap}
A continuous function $f:\R\to\C$ is called \textit{Besicovitch
almost periodic} if, for each $\eps>0$, there is a trigonometric
polynomial $P_{\eps}$ such that
\[
\|f-P_{\eps}\|^{}_{\mathcal{B}}<\eps \ts.
\]
\end{definition}
When  we ask if the two conditions in Theorem~\ref{thm:bohr} are also equivalent for the Besicovitch semi-norm, we are asking 
whether Besicovitch and mean almost periodicity are equivalent. It turns out that they are not.
Every Besicovitch almost periodic function is mean almost periodic. On the other hand, the models in
Example~\ref{ex1} and Example~\ref{ex2} give, after convolutions with continuous functions, examples of
functions that are mean almost periodic but not Besicovitch almost periodic.

Finally, let us introduce the last notion of almost periodicity that we want
to discuss in this article. In order to do so, we first define
\[
\|f\|^{}_{\mathcal{W}}:= \limsup_{n\to\infty}\, \sup_{t\in \R}\, \frac{1}{2n}
\int_{-n}^{n} |f(x-t)| \, \dd x    \ts.
\]
Once again, we do not obtain a norm but a semi-norm. For every
continuous function $f$ with compact support, one still has $\|f\|^{}_{\mathcal{W}} = 0$.

\begin{definition}\label{def:wap}
A continuous function $f:\R\to\C$ is called \textit{Weyl almost
periodic} if, for each $\eps>0$, there is a trigonometric polynomial
$P_{\eps}$ such that
\[
\|f-P_{\eps}\|^{}_{\mathcal{W}}<\eps \ts.
\]
\end{definition}

There is a hierarchy among the almost periodic functions that we
have discussed so far.

\begin{proposition} \label{prop:inclusions}
Bohr $\implies$ Weyl $\implies$ Besicovitch $\implies$ mean.
\end{proposition}

This hierarchy follows immediately from the following inequalities, which are easy to establish:
\[
\|f\|^{}_{\mathcal{B}} \leqslant\|f\|^{}_{\mathcal{W}} \leqslant \|f\|^{}_{\infty} \ts,
\]
see \cite{LSS} for details. None of the arrows in Proposition~\ref{prop:inclusions} can be reversed, as the examples covered above show.

\begin{remark}
Earlier, we defined limit-periodic and limit-quasiperiodic functions as limits of sequences of periodic and quasiperiodic functions, with respect to the supremum norm. They can also be defined using the Besicovitch or Weyl semi-norm instead. Next, we will construct a function which is limit-quasiperiodic but not limit-periodic with respect to the Weyl semi-norm. However, it is not Bohr almost periodic. Hence, it cannot be limit-quasiperiodic with respect to the supremum norm.
\end{remark}

\begin{example}
Let us consider the substitution
\[
L \mapsto LLSS \qquad \text{ and } \qquad S\mapsto LSS \ts,
\]
which leads to the bi-infinite word
\[
\ldots LSSLLSSLSSLSS|LLSSLLSSLSSLS\ldots
\]
Similar to the Fibonacci example, we turn every $S$ into a small interval of length $1$ and every $L$ into a long interval, this time, of length $\sqrt{2}$. So, the substitution rule inflates every interval by the factor $2+\sqrt{2}$. Next we put an isosceles triangle of height $1$ on every long interval, and an isosceles triangle of height $\tfrac{1}{2}$ on every small interval. Since the inflation factor $2+\sqrt{2}$ is an irrational number which is not a unit in $\Z[\sqrt{2}]$, the resulting function is limit-quasiperiodic but not limit-periodic with respect to the Weyl semi-norm, see \cite{GK}.
\end{example}

\section{The mathematical setup for
diffraction}\label{section-setup-diff}
The systematic setup for
mathematical diffraction theory for aperiodic sets goes back  to Hof
\cite{Hof}, compare de Bruijn \cite{deB-1,deB-2} for an earlier
treatment as well. A leisurely introduction to the topic with further
references can be found in \cite{L-PM,Cow}, compare \cite{BL-deformed,BL-Bremen}. For a monograph on the whole
field of mathematical treatment of quasicrystals we refer to
\cite{TAO}. Here, we discus the main ingredients.

Diffraction is considered in $d$-dimensional Euclidean space
$\RR^d$. To do the necessary averaging, we consider the sequence of
cubes $C_n = [-n,n]^d$.
A first model may start with a finite subset $F$ in $\RR^d$
which is thought of as modelling the positions of the atoms of the
piece of matter to be analyzed. Recall that we associate to this subset its
Dirac comb $\delta_{F} :=\sum_{x\in F} \delta_x$.
The diffraction then comes about as the square of the absolute value of the
\textit{Fourier transform} of the Dirac comb
\[\CalF
(\delta_{F})(y) := \sum_{x\in F} \e^{- 2\pi\im x\cdot
y} \ts.
\]
Hence, the intensity is given by
\[
I^{}_F(y)=|\CalF(\delta^{}_{F} ) (y)|^2
 = \sum_{x,t\in F} \e^{2\pi\im (x-t)\cdot y} \ts.
 \]
This approach can be summarized in a Wiener diagram, see Figure~\ref{fig:wiener}.

\begin{figure}[H]
\centering
\begin{tikzpicture}
  \matrix (m) [matrix of math nodes,row sep=3em,column sep=8em,minimum width=2em]
  {
\delta^{}_F & \delta^{}_F \ast \delta^{}_{-F}\\
     \mathcal{F}(\delta^{}_F) & {} |\mathcal{F}(\delta^{}_F)|^2 \\ };
  \path[-stealth]
    (m-1-1) edge node [left] {$\mathcal{F}$} (m-2-1)
            edge node [above] {$*$} (m-1-2)
    (m-2-1) edge node [below] {$| \cdot |^2$} (m-2-2)
    (m-1-2) edge node [right] {$\mathcal{F}$} (m-2-2);
\end{tikzpicture}
\caption{Wiener diagram for the diffraction of finite samples.}  \label{fig:wiener}
\end{figure}

We now turn to an infinite point set $\Lambda$ in $\R^d$. In this case, we have to consider the intensity per unit
volume, as the total intensity diverges. So, let us consider
the finite subset $\Lambda \cap C_n \mbox{
finite}$, for all $n$. Then, we define the intensity
\[
I_\Lambda  := \lim_{n\to \infty} \frac{1}{|C_n|}  \,
I_{\Lambda \cap C_n} \ts,
\]
where we assume that the limit exists. A
short computation then gives
\begin{eqnarray*}
I_\Lambda&=& \lim_{n\to \infty} \CalF \left( \frac{1}{|C_n|}
\delta_{\Lambda \cap C_n} \ast \delta_{-(\Lambda \cap C_n) } \right)\\[3mm]
&=& \CalF \left( \,\lim_{n\to \infty}
\frac{1}{|C_n|}\delta_{\Lambda
\cap C_n} \ast \delta_{-(\Lambda \cap C_n) }\right)\\
&=& \CalF (\gamma_\Lambda)
\end{eqnarray*}
(in the sense of measures) with
\[
 \displaystyle  \gamma_\Lambda = \lim_{n\to \infty}
\frac{1}{|C_n|} \delta_{\Lambda \cap C_n} \ast
\delta_{-(\Lambda \cap C_n)}  \ts.
\]

Let us now try to extend this definition to measures, to get a theory which also covers modeling by densities. 

Starting with a translation bounded  measure $\mu$, we form its
\textit{autocorrelation} (or averaged 2-point correlation)  $\gamma_\mu$ as
\[
\gamma_\mu
           :=\lim_{n\to\infty} \frac{\mu|_{C_n}\ast
             \widetilde{\mu|_{C_n}}}{|C_n|}  \ts.
\]
The  \textit{diffraction measure}
$\widehat{\gamma^{}_{\mu}}$ is the Fourier transform of autocorrelation, i.e.
$\CalF (\gamma^{}_{\mu})=\widehat{\gamma^{}_{\mu}}$.

\begin{figure*}
\centering
\begin{tikzpicture}
  \matrix (m) [matrix of math nodes,row sep=3em,column sep=8em,minimum width=2em]
  {
\underbrace{\delta^{}_{\mu}}_{\mbox{structure}} & \underbrace{\gamma^{}_{\mu}}_{\mbox{autocorrelation}}\\
 \underbrace{\sum_{y \in \points} A_y \delta_{y}}_{\mbox{Fourier--Bohr series}} & {}  \underbrace{\widehat{\gamma}=\sum_{y \in \points} |A_y|^2 \delta_{y}}_{\mbox{diffraction}} \\ };
  \path[-stealth]
    (m-1-1) edge node [left] {$\sim$} (m-2-1)
            edge node [above] {averaged convolution} (m-1-2)
    (m-2-1) edge node [below] {$| \cdot |^2$} (m-2-2)
    (m-1-2) edge node [right] {$\mathcal{F}$} (m-2-2);
\end{tikzpicture}
\caption{Wiener's diagram for Besicovitch almost periodic measures.} \label{fig:wiener-BAP}
\end{figure*}
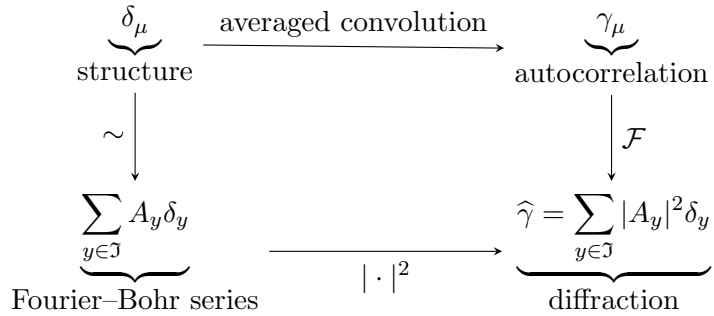

The preceding considerations yield the following
averaged version of one half of Wiener's diagram
\[
\mu
     \xrightarrow{\text{averaged convolution}}\gamma^{}_{\mu}
     \xrightarrow{\quad \mathcal{F}\quad} \widehat{\gamma^{}_{\mu}}
\]
Let us conclude by observing that, while in general the other half of the Wiener Diagram does not make sense, there is a natural way to make sense of it
in the case of Besicovitch almost periodic measures, see Figure~\ref{fig:wiener-BAP}.

\section*{Acknowledgements}
The authors would like to thank Michael Baake for
many suggestions and comments that greatly improved the quality of the paper. DL and TS would like to thank Ron Lifshitz for the invitation to a most inspiring workshop. NS was supported by the Natural Sciences and Engineering Council of Canada via grant 2020-00038, and he is grateful for the support. TS was supported by the German Research Foundation (DFG) via the CRC 1283.

\end{document}